\documentclass[pra,twocolumn,superscriptaddress,showpacs,amsmath,amssymb]{revtex4-1}
\usepackage[dvipdfmx]{graphicx,color}
\usepackage{subfigure}
\usepackage{natbib}
\usepackage{bm}
\usepackage{siunitx}
\usepackage{ulem}
\usepackage{lineno}

\def\U#1{{\rm #1}} 
\def\u#1{_{\rm #1}}
 
\newcommand{\od}[2]{\frac{\mathrm{d} #1}{\mathrm{d} #2}}
\def\gammaexp{\gamma\u{all}^{\U{(ex)}}}

\begin{document}
\title{
  Optical frequency tweezers 
}
\author{Rikizo Ikuta}
\affiliation{Graduate School of Engineering Science, Osaka University,
  Toyonaka, Osaka 560-8531, Japan}
\affiliation{
  Center for Quantum Information and Quantum Biology, 
  Osaka University, Osaka 560-8531, Japan}
\author{Masayo Yokota}
\affiliation{Graduate School of Engineering Science, Osaka University,
  Toyonaka, Osaka 560-8531, Japan}
\author{Toshiki Kobayashi}
\affiliation{Graduate School of Engineering Science, Osaka University,
  Toyonaka, Osaka 560-8531, Japan}
\affiliation{
Center for Quantum Information and Quantum Biology, 
Osaka University, Osaka 560-8531, Japan}
\author{Nobuyuki Imoto}
\affiliation{
Center for Quantum Information and Quantum Biology, 
Osaka University, Osaka 560-8531, Japan}
\author{Takashi Yamamoto}
\affiliation{Graduate School of Engineering Science, Osaka University,
Toyonaka, Osaka 560-8531, Japan}
\affiliation{
Center for Quantum Information and Quantum Biology, 
Osaka University, Osaka 560-8531, Japan}

\begin{abstract}
We show a concept of {\it optical frequency tweezers} 
for tweezing light in the optical frequency domain with a high resolution, 
which is the frequency version of the optical tweezers for the spatial manipulation of microscopic objects. 
We report the proof-of-principle experiment 
via frequency conversion inside a cavity only for the converted light. 
Owing to the atypical configuration, 
the experimental results 
successfully achieve the tweezing operation in the frequency domain, 
which picks a light at a target frequency
from the frequency-multiplexed input light and converts it to a different frequency, 
without touching any other light sitting in different frequency positions 
and shaking the frequency by the pump light. 
\end{abstract}
\maketitle

\section{introduction}
Optical tweezers~\cite{Ashkin1970,Ashkin1986}
that pick up and move microscopic objects using a highly focused laser 
are an indispensable technology that support modern science and engineering 
such as the optical trap of atoms~\cite{Chu1986}, 
manipulation of atoms in an optical lattice~\cite{Browaeys2020},
and biological science applications~\cite{Ashkin1987}. 
Similar to the optical tweezers for spatially deployed objects,
to manipulate photonic lattices in frequency synthetic dimension~\cite{Hu2020,Buddhiraju2021},
there is an increasing demand for tweezing light
in the optical frequency domain with a high resolution 
that precisely picks up only light in a target frequency mode 
from broadened or densely frequency multiplexed light 
and converts it to any frequency mode. 
In addition, it is also required in the field of quantum information processing 
for realizing photonic quantum computation 
based on quantum frequency combs~\cite{Kues2017,Lukens2017,Niu2018,Kues2019,Reimer2019}, 
and quantum internet~\cite{Kimble2008} to interconnect various quantum systems 
including frequency/wavelength-division multiplexing technologies~\cite{Sinclair2014,Wengerowsky2018,Lingaraju2021}. 
However, such {\it optical frequency tweezers} have never been realized. 

In this study, 
we propose an optical system for the optical frequency tweezers, 
as illustrated in Fig.~\ref{fig:concept}~(a), 
and perform its proof-of-principle demonstration. 
To this end,
we adopt a frequency conversion based on a nonlinear optical interaction 
inside a cavity that only confines light to be generated by the conversion process. 
Fig.~\ref{fig:concept}~(b) presents the system design 
based on a second-order nonlinearity~($\chi^{(2)}$). 
The input signal interacts with the cavity mode via the frequency conversion 
with a strong pump light, 
and then the converted light is extracted because of the cavity decay. 
The system is considered as an all optical implementation of 
a multiplexed $\Lambda$-type three-level system, 
as illustrated in Fig.~\ref{fig:concept}~(c). 
The cavity resonant frequency modes adopted as the discretized excited levels 
modulate the bandwidth of the nonlinear optical interaction 
while possessing broadband bandwidths for the signal and pump modes. 
Therefore, it is feasible for the pump light to select a target frequency mode 
to be converted from a frequency-multiplexed input signal. 
It represents the minimum of required properties for the optical frequency tweezers.
In addition, 
the tweezers should precisely pick up the target without shaking 
and should not disturb the unconverted signals around the target. 
The unconventional cavity configuration addresses these requirements 
for the tweezers, which have never been reported 
with the use of multiply resonant cavities such as microresonator systems. 
There is no thermal fluctuation owing to no confinement of the pump light, 
and there is no disturbance and cavity loss of the unconverted light 
due to no confinement of the input signals. 

\section{basic concept}
\begin{figure*}
 \begin{center}
         \scalebox{1}{\includegraphics{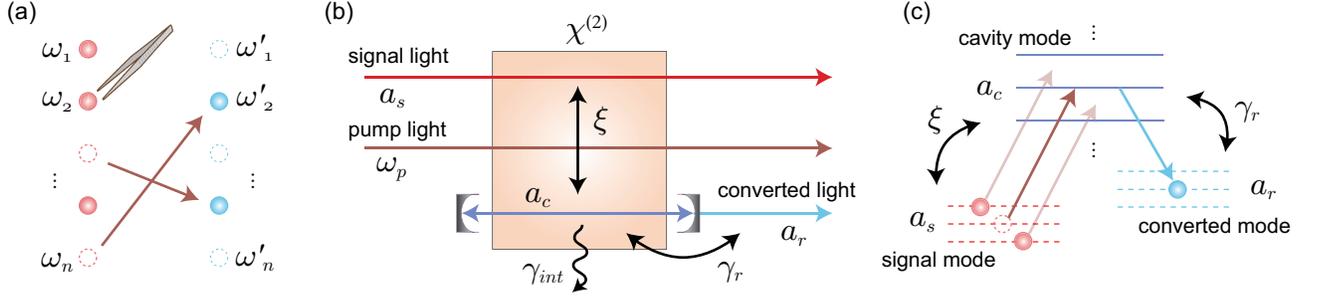}}
    \caption{
      (a) 
      Concept of optical frequency tweezers. 
      The tweezers pick up only light at the target frequency 
      and deploy it to the desired frequency mode. 
      (b) 
      Implementation of the optical frequency tweezers
      based on $\chi^{(2)}$-based frequency conversion 
      inside a cavity only for the converted light. 
      The mode coupling relationship is also illustrated. 
      $a_s$, $a_r$, and $a_c$ correspond to the signal, converted and cavity modes.
      $\omega_p$ denotes the pump frequency. 
      (c) 
      Diagram of the frequency tweezing system 
      when only one of the converted frequency of the input frequencies 
      is resonant on the cavity. 
      Only the middle of the input signal can be converted. 
    }
 \label{fig:concept}
 \end{center}
\end{figure*}
The optical frequency tweezers introduced in this study 
are realized by the photonic $\Lambda$-type three-level system
based on the $\chi^{(2)}$ nonlinearity 
inside a single-sided cavity around the converted frequencies~(Fig.~\ref{fig:concept}~(b)). 
We focus on the frequency conversion process around the resonant frequency 
of the cavity described by the single mode $a_c$. 
The cavity mode is coupled to two modes outside the cavity, namely $a_r$ and $a_s$, 
of which the former corresponds to the cavity decay through its single side 
and the latter corresponds to the input signal field obtained via the $\chi^{(2)}$-based frequency conversion. 
The interaction Hamiltonians of these coupling systems are described by 
$H_C=i\sqrt{\gamma_r} a_ca_r^\dagger + H.c.$ and $H_{NL}=i\xi a_ca_s^\dagger + H.c.$, 
where $H.c.$ denotes the Hermitian conjugate, 
$\sqrt{\gamma_r}$ is a coupling constant between the internal and external modes of the cavity, 
and $\xi=|\xi|e^{i\phi}$ represents an effective coupling constant of the frequency conversion 
that is proportional to the complex amplitude of the sufficiently strong pump light
with its phase $\phi$. 
From the system Hamiltonian $H_C+H_{NL}$, 
the time evolution of the cavity mode $a_c = a_c(t)$ 
in a frame rotating at the resonant frequency $\omega_c$ is described by~\cite{Collett1984}
\begin{eqnarray}
    \od{a_c}{t} = i\Delta_ca_c -\frac{\gamma\u{all}+|\xi|^2}{2}a_{c}
  + \sqrt{\gamma_r} a_{r, IN} + \xi^* a_{s,IN}, 
  \label{eq:ac}
\end{eqnarray}
where $\gamma\u{all} := \gamma_r + \gamma\u{int}$ is the total loss of the cavity 
including the internal loss $\gamma\u{int}$. 
$\Delta_c := \omega - \omega_c$ denotes the detuning of the light in the cavity, 
where $\omega$ represents the frequency of mode $a_c$, 
and is determined by the frequencies $\omega_s$ and $\omega_p(=\omega_s-\omega)$ 
of mode $a_s$ and the pump light.
$a_{s,IN}$ is the input signal field to be converted. 
$a_{r,IN}$ is the external drive field,
and is assumed to be the vacuum for the standard usage of this device, 
which we will describe in the next paragraph. 
This drive field can be used to control the frequency conversion process 
of $a_{s,IN}$, which will be explained in detail based on Eq.~(\ref{eq:eit}).
The system described by Eq.~(\ref{eq:ac})
can be regarded as the $\Lambda$-type three-level quantum system, 
which comprises the excited level $a_c$ and two ground levels, $a_s$ and $a_r$, 
as illustrated in Fig.~\ref{fig:concept}~(c). 
We note that this diagram does not present the energy relationship among the relevant modes 
as conventional ones in $\Lambda$-type atomic systems. 
The excited level $a_c$ in the diagram exhibits the energy lower than that of the ground level $a_s$, 
while $a_s$ is excited to $a_c$ by the coupling based on the frequency conversion. 
In addition, although $a_c$ and another ground level $a_r$ exhibit the same energy, 
$a_c$ decays to $a_r$ due to the spatial coupling at the end mirror. 

The frequency conversion from $a_s$ to $a_r$ is the case of $a_{r,IN}=0$. 
From Eq.~(\ref{eq:ac}) and the input-output relationships~\cite{Walls2007} 
$a_{s,IN} - a_{s,OUT} = \xi a_c$ and
$a_{r,IN} + a_{r,OUT} = \sqrt{\gamma_r} a_c $, 
we obtain 
\begin{eqnarray}
  t_{ss} := \frac{a_{s,OUT}}{a_{s,IN}} 
            &=& \frac{\frac{1}{2}(1-\widetilde{C})-i\widetilde{\Delta}_c}
                {\frac{1}{2}(1 + \widetilde{C})-i\widetilde{\Delta}_c},
                \label{eq:t}\\
  r_{rs} := \frac{a_{r,OUT}}{a_{s,IN}} &=& \sqrt{\widetilde{\gamma}_r}
                \frac{e^{-i\phi}\sqrt{\widetilde{C}}}
                {\frac{1}{2}(1 + \widetilde{C})-i\widetilde{\Delta}_c}, 
  \label{eq:r}
\end{eqnarray}
where $\widetilde{\gamma}_r:=\gamma_r/\gamma\u{all}$
and $\widetilde{\Delta}_c:=\Delta_c/\gamma\u{all}$. 
$\widetilde{C}:=|\xi|^2/\gamma\u{all}$ corresponds to the cooperativity parameter. 
$t_{ss}$ and $r_{rs}$ are complex amplitudes 
for the staying probability $T(=|t_{ss}|^2)$ at mode $a_s$
and the transition probability $R(=|r_{rs}|^2)$ to mode $a_r$. 
These equations hold at each frequency interval
corresponding to the free spectral range~(FSR) of the cavity
with the same level of the nonlinearity strength, 
regardless of the dispersion of the medium. 
This is because the singly resonant cavity structure eliminates
the restriction in conventional doubly or triply resonant cavities, 
such that the relevant three frequencies satisfying the energy conservation 
must be simultaneously resonant on the cavities~\cite{Ikuta2019}. 

From the analysis, 
we can observe remarkable properties in the frequency conversion 
based on the $\Lambda$-type structure as follows: 
(i)
The device can pick up any single frequency mode 
from multiplexed input frequency modes, 
and then convert to any frequency mode by designing the FSR and pump frequency, 
such that only the
converted frequency corresponding to the target frequency to be chosen for conversion 
is resonant, as illustrated in Fig.~\ref{fig:concept}~(c). 
(ii)
The remaining signal light which is far from the cavity resonances 
passes through the device without any disturbance and loss, 
which is observed from $T = 1$ for sufficiently large $\widetilde{\Delta}_c$ in Eq.~(\ref{eq:t}). 
(iii)
No frequency shake occurs during the pick up process. 
This is due to the absence of the thermal heating effect
triggered by the pump light, which is not confined in the cavity, 
as well as the stable operation without a severe frequency locking. 
Properties (i) -- (iii) are exactly the requirements for the ``tweezers.'' 
Hence, we call the device optical frequency tweezers. 

In addition to the aforementioned basic properties,
our device also has advantages as the cavity-based frequency converter as follows: 
(iv)
It has no optical impedance matching problem for the input signal, 
regardless of the cavity enhancement characterized by $\widetilde{C}$. 
(v)
The spectral shape of the converted light is not split 
or distorted, and obeys the Lorentzian, 
which differs from the mode splitting phenomena
analogous to the Autler-Townes splitting that
appears in coupled resonator systems~\cite{Peng2014,Guo2016}. 
(vi)
The maximum conversion efficiency $\widetilde{\gamma}_r$ would be higher
than that determined by the product of cavity losses for both signal and converted modes 
in the frequency conversion based on multiply resonant cavities~\cite{Li2016,Guo2016}. 
(vii)
All optical adjustments of $\widetilde{\Delta}_c$ are possible by the pump frequency tuning. 

\begin{figure*}[t]
 \begin{center}
   \scalebox{1}{\includegraphics{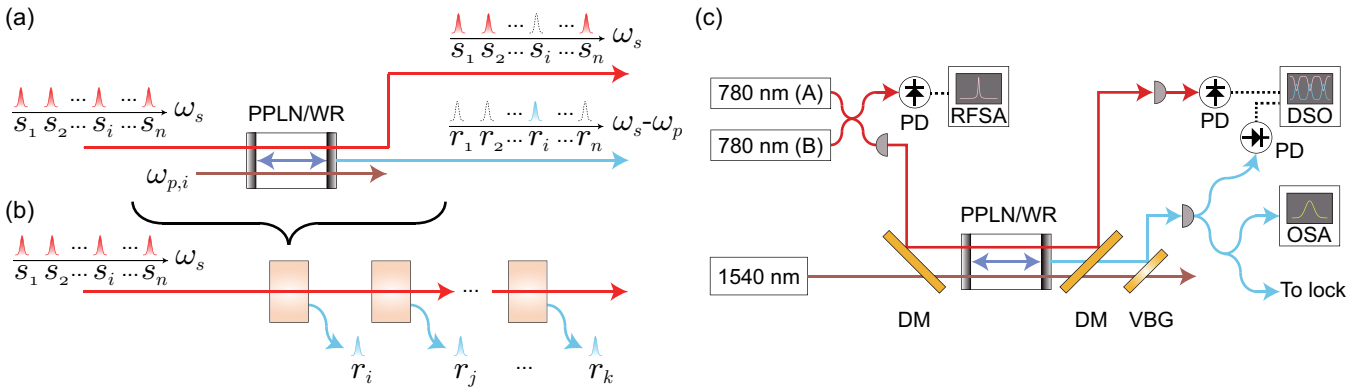}}
    \caption{
      (a) 
      Setup for picking a frequency mode around $s_i$ 
      and deploying it to another mode around $r_i$ using a pump light at $\omega_{p,i}$. 
      The $\chi^{(2)}$ medium for the frequency conversion is the PPLN/WG. 
      (b) 
      Cascade use of the above setup 
      for any-to-any tweezing, as presented in Fig.~\ref{fig:concept}~(a). 
      (c) 
      Experimental setup for the proof-of-principle experiment. 
      The signal light A at \SI{780}{nm} is used to characterize
      the performance of the frequency conversion.
      The demonstration of the tweezing operation is performed 
      by using both signal lights A and B
      with frequency detuning $\nu_s$.
      PD; photodetector,
      RFSA; radio frequency spectral analyzer, 
      DSO; digital sampling oscilloscope, 
      OSA; optical spectral analyzer. 
    }
 \label{fig:setup}
 \end{center}
\end{figure*}
\section{experiments}
The experimental design for the optical frequency tweezers 
is presented in Fig.~\ref{fig:setup}~(a). 
The scenario considered here involves tweezing one tooth 
from a frequency comb or quantum frequency comb~\cite{Kues2019}. 
When all the frequency intervals and the cavity FSR 
satisfy a relationship in which only 
the converted frequency corresponding to the target tooth 
is resonant on the cavity, 
the tweezers can pick up and convert the frequency of the tooth 
and leave the rest of the teeth on the unconverted modes without disturbance. 
Consequently, the cascade use of the tweezers presented in Fig.~\ref{fig:setup}~(b) 
can realize any-to-any tweezing operation, as illustrated in Fig.~\ref{fig:concept}~(a). 

The experimental setup for the proof-of-principle demonstration 
is shown in Fig.~\ref{fig:setup}~(c). 
The continuous wave signal light around \SI{780}{nm}~(corresponding to $\omega_s$) with $\sim$~\SI{1}{mW} 
and the pump light for the tweezers at \SI{1540}{nm}~($\omega_p$) 
are combined at a dichroic mirror~(DM).
Subsequently, they are focused on the
periodically poled lithium niobate waveguide resonator~(PPLN/WR), 
which only confines the converted light around \SI{1581}{nm}~($\omega_r$).

PPLN/WR satisfies the type-0 quasi-phase-matching condition 
for the vertically polarized light. 
The length of the waveguide is \SI{20}{mm}.
The bandwidth of frequency conversion for the input light at \SI{780}{nm} 
is over \SI{150}{GHz}~\cite{Ikuta2021}. 
The FSR of the resonator is $\nu\u{FSR} = \SI{3.5}{GHz}$~\cite{Ikuta2018-2}. 
In addition, the dielectric multilayers on the end faces of the waveguide 
achieve high reflectance of $\sim$ \SI{94}{\%} for light around \SI{1581}{nm}. 
We determined the full width at the half maximum~(FWHM) of the cavity
without the internal loss by the reflectance as $\gamma_0:=\SI{71}{MHz}$. 
The reflectance for \SI{780}{nm} is a few percent and thus the anti-reflection coating is achieved. 
The transmittance of \SI{780}{nm} light to the PPLN/WR is \SI{90}{\%}. 
Although the reflectance for \SI{1540}{nm} is slightly larger at approximately \SI{30}{\%}, 
it forms a very lossy cavity. 
Considering the fineness is smaller than $\pi$~\cite{Rogener1988,Stefszky2018},
there should not be any cavity enhancement effect for the frequency conversion. 

After the PPLN/WR, 
the signal light and converted light are extracted using another DM 
and a volume Bragg grating~(VBG)
with its center wavelength and bandwidth of \SI{1581}{nm} and \SI{1}{nm}, respectively. 
They are coupled to single mode fibers 
followed by proper experimental apparatuses. 

\begin{figure*}[t]
 \begin{center}
   \scalebox{1}{\includegraphics{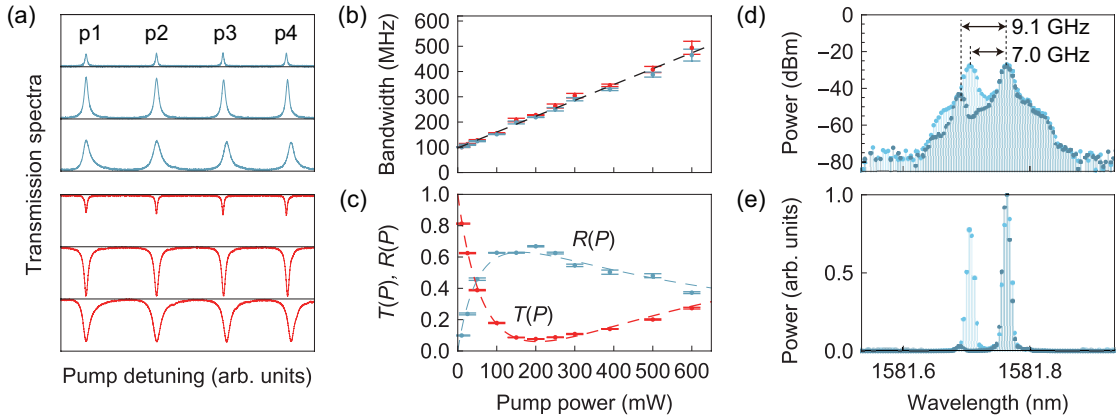}}
   \caption{
     (a) 
     Observed four transmission spectra~(p1 -- p4)
     related to the longitudinal cavity modes separated by the FSR
     of the 1581-nm~(upper) and 780-nm light~(lower) 
     for pump power \SI{25}{mW}, \SI{200}{mW}, and \SI{500}{mW} from the top. 
     (b)
     Pump power dependency of 
     observed bandwidths for 780-nm~(red) and 1581-nm light~(blue).
     (c)
     Pump power dependency of 
     the transition probability $R(P)$~(blue) and staying probability $T(P)$~(red). 
     (d)
     Observed spectra around \SI{1581}{nm} for 
     $\nu_s = \SI{7.0}{GHz} = 2\,\nu\u{FSR}$~(light blue)
     and
     $\nu_s = \SI{9.1}{GHz} = 2.6\,\nu\u{FSR}$~(dark blue).
     (e)
     Same figure as Fig.~\ref{fig:result}~(d)
     with the linear scale for the vertical axis. 
   }
 \label{fig:result}
 \end{center}
\end{figure*}
First, we characterize the performance of the frequency conversion 
from \SI{780}{nm} to \SI{1581}{nm}. 
We adopt the signal light emerging from light source~A. 
By scanning the pump frequency in the range of approximately \SI{15}{GHz}, 
we observed
four peaks~(labelled p1 -- p4) in the transmission spectra
for \SI{780}{nm} and \SI{1581}{nm}, 
which is related to the longitudinal cavity modes separated by the FSR. 
The spectra are presented in Fig.~\ref{fig:result}~(a). 
As an example, we focus on the rightmost peaks~(p4) of the figure. 
The bandwidths of the frequency conversion around the peaks
are presented in Fig.~\ref{fig:result}~(b). 
As expected in Eqs.~(\ref{eq:t}) and (\ref{eq:r}), 
the bandwidths obtained by the 780-nm and 1581-nm light 
are almost the same and proportional to the pump power $P$ measured before the PPLN/WR. 
The best fit to the data with a function $\alpha P + \gammaexp$ 
gives $\alpha =\SI{0.68}{MHz/mW}$ and $\gammaexp=\SI{106}{MHz}$. 
$\gammaexp$ reflects the total loss of the resonator 
corresponding to $\gamma\u{all}$ introduced in Eq.~(\ref{eq:ac}), 
which includes the internal loss of the resonator, as well as the loss from the end mirrors. 

For the conversion efficiency,
its maximum is limited to
$\widetilde{\gamma_r}=\gamma_0/\gammaexp=\SI{67}{\%}$ 
from Eq.~(\ref{eq:r}) 
for the internal loss of the resonator 
estimated to be $\gamma\u{int}=\SI{35}{MHz}$.  
Besides, a mode mismatch of the signal and the converted modes 
propagating in the waveguide should degrade the conversion efficiency. 
To estimate the amount $M$ of the mode match, 
we plot the bottom peaks of the observed spectra for the780-nm light 
normalized by the value at $P=\SI{0}{mW}$ in Fig.~\ref{fig:result}~(c). 
The best fit to the data with a function $M T(P) + (1-M)$ gives $M=0.94$, 
where $T(P)=|t|^2$ is given by Eq.~(\ref{eq:t}) with $\widetilde{\Delta}_c=0$. 
Accordingly, the maximum of the achievable internal conversion efficiency is 
estimated as $\eta\u{max}:=M\widetilde{\gamma_r}=0.63$. 
By normalizing the observed peaks for the 1581-nm light with the maximum, 
we can observe the internal conversion efficiency, as presented in Fig.~\ref{fig:result}~(c). 
From the best fit to the data with a function 
$\eta\u{max}R(P)$ given by Eq.~(\ref{eq:r}) with $\widetilde{\Delta}_c=0$, 
the unit power cooperativity is estimated to be $\beta=\SI{0.0062}{mW^{-1}}$, 
which agrees well with $\alpha/\gammaexp=\SI{0.0064}{mW^{-1}}$, 
as estimated from the experimental result about the bandwidth. 
Although the PPLN/WR used in this experiment is the Fabry-P\'{e}rot cavity 
and the converted light emerges from its two end faces, 
an asymmetric mirror coating could limit the output to one side only, 
or these two modes can be coherently combined by adopting an interferometer 
such as the Sagnac loop~\cite{Yamazaki2021}.

We notice that
the pump power for the maximum conversion efficiency~($\widetilde{C}=1$) 
is $\beta^{-1}\sim\SI{160}{mW}$ 
while the maximum in the case without cavities was \SI{700}{mW}~\cite{Ikuta2011}. 
This clearly indicates the cavity enhancement effect
without the pump light confinement. 
In addition, the good agreement of the theoretical curves 
and the experimental results for all pump powers implies 
the absence of unwanted nonlinear optical interaction 
that facilitates noise light generation around the target wavelengths to be converted 
and the pump power consumption
observed in the $\chi^{(2)}$-based frequency conversion
with the pump light confinement~\cite{Ikuta2021}. 

We performed the same analysis as described above for the other peaks in Fig.~\ref{fig:result}~(a).
The estimated values are listed in Table~\ref{tbl:ex_tbl}. 
It is evident that almost the same values are obtained for every parameter. 
This implies that this device achieved 
equal performance on the frequency conversion around every resonant frequency. 
\begin{table}[t]
\begin{center}
\begin{tabular}
  {|c|cccc|}
  \hline
  peak number & p1 & p2 & p3 & p4 \\ \hline\hline
  $\alpha$~(MHz/mW) & $0.63\pm 0.03$ & $0.74\pm 0.02$ & $0.63\pm 0.02$ & $0.68\pm 0.02$ \\ \hline
  $\gammaexp$~(MHz) & $117\pm 4$ & $ 105 \pm 3$ & $106 \pm 2$ & $106 \pm 2$ \\ \hline
  $M$~(\%) & $93\pm 0.2$ & $ 94 \pm 0.1$ & $94 \pm 0.2$ & $94 \pm 0.1$ \\ \hline
  $10^3 \beta$~(/mW) & $5.6\pm 0.1$ & $ 5.9 \pm 0.1$ & $6.1 \pm 0.1$ & $6.2 \pm 0.1$ \\ \hline
\end{tabular}
 \caption{
   Estimated values on the four peaks. 
 \label{tbl:ex_tbl}}
 \end{center}
\end{table}

Next, we perform the tweezing operation. 
To suppress the shaking during the picking and maximize the conversion efficiency, 
we adopted part of the converted light to lock the pump frequency. 
The observed power of signal A after PPLN/WR was reduced from \SI{0.6}{mW} to \SI{0.04}{mW}
via frequency conversion with the frequency-locked pump light. 
This decrease ratio of 0.93 agrees well with the observed value $M$ in the previous section. 
Under this setup, we input an additional signal light denoted by B 
with a frequency difference $\nu_s$ from the frequency of signal A. 
We show the observed spectra of the converted light around \SI{1581}{nm}
in Figs.~\ref{fig:result}~(d) and (e) 
with logarithmic and linear scales for the vertical axes, respectively. 
It can be observed that for $\nu_s = \SI{7.0}{GHz} = 2\,\nu\u{FSR}$, 
both signals were simultaneously converted to \SI{1581}{nm}. 
In contrast, for $\nu_s = \SI{9.1}{GHz} = 2.6\,\nu\u{FSR}$, 
the converted light by signal B was
significantly smaller~(a suppression of \SI{16}{dB})
because the $\widetilde{\Delta}_c$ value for signal B is sufficiently large. 
Based on this result, 
we conclude that the proof-of-principle experiment
for the optical frequency tweezers was successfully achieved. 

\begin{figure*}
 \begin{center}
   \scalebox{1}{\includegraphics{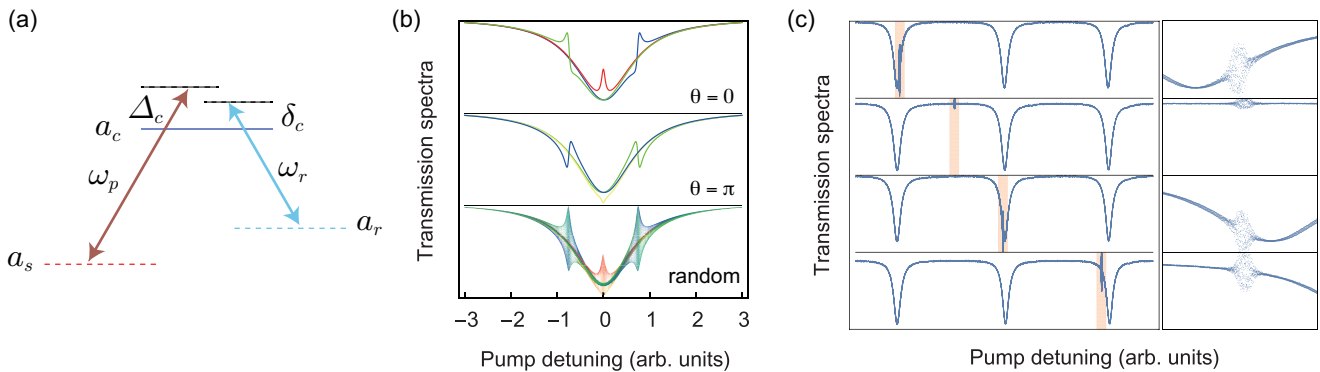}}
    \caption{
      (a) 
      Diagram of the three-level system with an external input light from mode $a_r$. 
      (b)
      Theoretical simulation of the interference effect 
      by using Eq.~(\ref{eq:eit}) with
      $\phi=0$ and $\widetilde{C}\widetilde{\gamma_r}=0.25$.
      The control light is assumed to be 
      $a_{r,IN}/a_{s,IN}=e^{i\theta} 0.5\widetilde{\Gamma}/(\widetilde{\Gamma}+i\widetilde{\delta}_c)$
      with $\widetilde{\Gamma}=0.05$.
      The top figure presents three curves 
        for $\widetilde{\delta}_c=\widetilde{\Delta}_c - 0.75$~(green), 
        $\widetilde{\Delta}_c$~(red), and 
        $\widetilde{\Delta}_c + 0.75$~(blue) 
        in the case of $\theta=0$. 
        The middle figure shows three curves 
        for 
        $\widetilde{\delta}_c=\widetilde{\Delta}_c - 0.75$~(blue), 
        $\widetilde{\Delta}_c$~(yellow), and 
        $\widetilde{\Delta}_c + 0.75$~(green)
        in the case of $\theta=\pi$.
        The bottom figure shows the curves 
        for the three detunings
        $\widetilde{\delta}_c=\widetilde{\Delta}_c \pm 0.75, \widetilde{\Delta}_c$, 
        each with the random phase of 
        $\theta=k\pi/20$~($k=0,\ldots, 40$). 
      (c)
      The observed transmission spectra at \SI{780}{nm}
      with the control light around \SI{1581}{nm} with four different detunings.
      The figures on the right are enlarged figures on the left. 
    }
 \label{fig:eit}
 \end{center}
\end{figure*}
The tweezing operation presented above is based on the frequency conversion of the 780-nm light. 
In addition to the function, 
the $\Lambda$-type system enables us to adopt an external field at \SI{1581}{nm}, 
to control the tweezing operation~(refer to Fig.~\ref{fig:eit}~(a)). 
The external field is described by $a_{r,IN}$ in Eq.~(\ref{eq:ac}),
and the complex amplitude of the staying probability becomes
\begin{eqnarray}
  \frac{a_{s,OUT}}{a_{s,IN}} 
  = \frac{e^{i\phi}\sqrt{\widetilde{C}\widetilde{\gamma}_r}a_{r,IN}/a_{s,IN}+\frac{1}{2}(1-\widetilde{C})-i\widetilde{\Delta}_c}
  {\frac{1}{2}(1 + \widetilde{C})-i\widetilde{\Delta}_c}.
  \label{eq:eit}
\end{eqnarray}
The equation above demonstrates that the output of the signal 
includes an interference effect between modes $a_r$ and $a_s$. 
Unlike the coupled-resonator-based $\Lambda$-type optical systems~\cite{Peng2014,Guo2016},
it is possible to freely tune the relative phase of the free fields. 
Consequently, both destructive and constructive interferences 
corresponding to the electromagnetically-induced transparency~(EIT)
and the electromagnetically-induced absorption~(EIA) based on the Fano interference
in atomic systems are achieved
without altering the cooperativity $\widetilde{C}$~\cite{Naweed2005}.
The theoretical simulation of the optical implementations of EIT and the EIA 
under an assumption that the frequency of the control light is locked 
to that of the pump light is presented in Fig.~\ref{fig:eit}~(b).
The results indicate the possibilities that EIT and the EIA 
can, respectively, obstruct and facilitate the tweezing operation 
close to the resonant frequencies. 
This property enables the device to provide additional functions. 
For example, 
the switching operation to all frequency modes by the pump light is performed at once 
for the frequency multiplexed input with the frequency interval corresponding to the FSR.
However, the control light enables us to switch one of the frequency modes. 
Another usage of this property involves performing the tweezing operation of the signal 
sensitive to the phase of the control beam as an additional reference light. 

The observed spectra of the 780-nm light for various wavelengths 
of the control light are presented in Fig.~\ref{fig:eit}(c). 
In response to the wavelength of the control light changes, 
the wavelength for observing the interference changed. 
As is expected, both the destructive and constructive interferences 
related to the EIT and EIA 
are observed at the control frequencies near the resonant points, 
while the interference signatures reflect the random phase structure 
similar to the bottom of Fig.~\ref{fig:eit}~(b) 
because the signal and control lasers were not phase locked. 
This result implies that the tweezing operation can be switched 
by using a more stable laser for the control field. 

\section{conclusion}
In this study, we presented the optical frequency tweezers 
with experimental implementation 
based on frequency conversion with the atypical cavity structure. 
It facilitated the precise frequency tweezing operation 
without shaking the frequency or disturbing the light, 
other than the target to be chosen for tweezing. 
Using the control light as the analogies of EIT and EIA in atomic systems, 
we demonstrated the possibilities of
suppressing and enhancing the tweezing operation.
Such optical frequency tweezers will be indispensable
as the channel exchange for a massive any-to-any channel switch 
among frequency-multiplexed channels,
which has the same role as a conventional channel exchange
for an any-to-any switch among spatially assigned channels. 
In addition to the above tweezing operation, 
we verified the advantages of our frequency converter. 
Importantly, the cavity enhancement without severe impedance matching 
and the higher maximum conversion efficiency will be beneficial 
in the frequency conversion of a single photon 
with quantum information that is never amplified. 

\section{acknowledgments}
R.I., N.I., and T.Y. acknowledge members of Quantum Internet Task Force (QITF)
for comprehensive and interdisciplinary discussions of the quantum internet.
This work was supported by Moonshot R \& D, JST JPMJMS2066; 
CREST, JST JPMJCR1671;
MEXT/JSPS KAKENHI Grants No. JP21H04445, JP20H01839,
and Asahi Glass Foundation.

\end{document}